\documentclass[11pt,a4paper]{article}
\usepackage[english]{babel}
\usepackage{amsmath,amsthm,amssymb,epsfig,latexsym}
\usepackage{color}
\usepackage{ulem}

\newcommand{\la}{u}
\newcommand{\muu}{v}
\newcommand{\lac}{u^{\scriptscriptstyle C}}
\newcommand{\lab}{u^{\scriptscriptstyle B}}
\newcommand{\muc}{v^{\scriptscriptstyle C}}
\newcommand{\mub}{v^{\scriptscriptstyle B}}
\newcommand{\as}{\lambda}
\newcommand{\bla}{\bar u}
\newcommand{\bmu}{\bar v}
\newcommand{\blac}{\bar{u}^{\scriptscriptstyle C}}
\newcommand{\blab}{\bar{u}^{\scriptscriptstyle B}}
\newcommand{\bmuc}{\bar{v}^{\scriptscriptstyle C}}
\newcommand{\bmub}{\bar{v}^{\scriptscriptstyle B}}

\newcommand{\Nmod}{\widehat{\mathcal{N}}^{(s)}}
\newcommand{\Ngltwo}{n}

\newcommand{\Not}{\mathcal{N}}
\newcommand{\Hot}{\mathcal{H}}

\newcommand{\Ntt}{\widetilde{\mathcal{N}}}
\newcommand{\Htt}{\widetilde{\mathcal{H}}}

\newcommand{\be}[1]{\begin{equation}\label{#1}}
\newcommand{\ba}[1]{\begin{multline}\label{#1}}
\newcommand{\ee}{\end{equation}}
\newcommand{\ea}{\end{eqnarray}}

\newcommand{\num}{\\\rule{0pt}{20pt}}

\newcommand{\dis}{\displaystyle}

\newcommand{\diag}{\mathop{\rm diag}}
\newcommand{\tr}{\mathop{\rm tr}}

\newtheorem{prop}{Proposition}[section]

 \makeatletter
 \@addtoreset{equation}{section}
 \makeatother
 
\newcommand{\bea}{\begin{eqnarray}}
\newcommand{\eea}{\end{eqnarray}}

\begin{document}

\begin{flushright}
LAPTH-045/14
\end{flushright}

\vspace{20pt}

\begin{center}
\begin{LARGE}
{\bf Determinant representations for\\ form factors
in quantum integrable models\\
 with $GL(3)$-invariant $R$-matrix }
\end{LARGE}

\vspace{40pt}

\begin{large}
{S.~Pakuliak${}^a$, E.~Ragoucy${}^b$, N.~A.~Slavnov${}^c$\footnote{pakuliak@theor.jinr.ru, eric.ragoucy@lapth.cnrs.fr, nslavnov@mi.ras.ru}}
\end{large}

 \vspace{12mm}

\vspace{4mm}

${}^a$ {\it Laboratory of Theoretical Physics, JINR, 141980 Dubna, Moscow reg., Russia,\\
Moscow Institute of Physics and Technology, 141700, Dolgoprudny, Moscow reg., Russia,\\
Institute of Theoretical and Experimental Physics, 117259 Moscow, Russia}

\vspace{4mm}

${}^b$ {\it Laboratoire de Physique Th\'eorique LAPTH, CNRS and Universit\'e de Savoie,\\
BP 110, 74941 Annecy-le-Vieux Cedex, France}

\vspace{4mm}

${}^c$ {\it Steklov Mathematical Institute,
Moscow, Russia}

\end{center}


\vspace{4mm}


\begin{abstract}
We obtain determinant representations for the form factors of the
monodromy matrix  entries in quantum integrable models solvable by the nested algebraic Bethe ansatz and possessing
$GL(3)$-invariant $R$-matrix. These representations can be used for the calculation of
correlation functions in the models of physical interest.
\end{abstract}

\vspace{1cm}


\vspace{2mm}

\section{Introduction}

The form factor approach is one of the most effective methods for calculating
correlation functions of quantum integrable models. Therefore, finding explicit and compact representations for
the form factors is an important task.
Currently there are several  methods to study  form factors of integrable systems. One of the first to be developed was the
so called `form factor bootstrap approach', which has been successfully applied to integrable
quantum field theory \cite{KarW78,Smi92b,CarM90,Mus92,FriMS90,KouM93,AhnDM93}. This method
is closely related to the one based on the conformal field theory and its perturbation \cite{Zam91,LukZ97,Luk99,LukZ01}.
It is also worth mentioning  the approach developed in \cite{JimMMN92,JimM95L,JimM96}, where
the form factors were studied via the representation theory of quantum
affine algebras. All the methods listed above
deal with quantum integrable models in infinite volume. Form factors in the models of finite volume
were studied in \cite{KojKS97,KitMT99} by the algebraic Bethe ansatz \cite{FadST79,FadT79,BogIK93L,FadLH96}.  In particular, this method was found to be very efficient for quantum spin chain models, for which the solution of the
quantum inverse scattering problem is known \cite{KitMT99,MaiT00}. Determinant representations for
form factors obtained in this framework were successfully used for the calculation of correlation functions \cite{KitKMST11,KitKMST12,CauM05,CauPS07}.

The results listed above mostly concern the models based on $GL(2)$ symmetry or its $q$-deformation.
Models with a higher rank symmetry were much less studied. At the same time such
models  play an important role in various applications. For instance, integrability has proved to be a very efficient tool for the calculation of scattering amplitudes in super-Yang-Mills theories. The calculation of these amplitudes  can be related to scalar products of Bethe vectors. In particular, in the $SU(3)$ subsector of the theory, one just needs the $SU(3)$ Bethe vectors. Hence, the knowledge of the form factors is very essential in this context.

Form factors of integrable models with symmetries of high rank also appear in
condensed matter physics, in particular in two-component Bose (or Fermi) gas and in the study of models of cold atoms (for e.g. ferromagnetism or phase separation). One can also mention 2-band Hubbard models (mostly in the half-filled regime), in the context of strongly correlated electronic systems. In that case, the symmetry increases when spin and orbital degrees of freedom are supposed to play a symmetrical role, leading to an $SU(4)$ or even an $SO(8)$ symmetry (see e.g. \cite{su4-Hub,so8-Hub}). All these studies require to look for integrable
models with $SU(N)$ symmetry, the first step being the $SU(3)$ case. In this context it is worth mentioning the work \cite{PozOK12}, where the form factors in the model of two-component Bose gas were studied.

In this article we give determinant representations
for  form factors in $GL(3)$-invariant quantum integrable models
solvable  by the nested algebraic Bethe ansatz \cite{KulRes83,KulRes81,KulRes82}. More precisely,
we calculate matrix elements of the monodromy matrix entries $T_{ij}(z)$  between on-shell Bethe vectors
(that is, the eigenstates of the transfer matrix). The determinant representations given in this paper are based
on the formulas obtained in \cite{BelPRS12b,BelPRS13a,PakRS14b}. There, however, we had slightly different representations for the form factors of the diagonal entries $T_{ii}(z)$ and the ones for $T_{ij}(z)$ with $|i-j|=1$. Furthermore, in the case of
the operators $T_{ii}(z)$ one had to consider two different cases depending on whether two Bethe vectors coincided or
were different. In this paper we give more uniform determinant representations for all form factors. We also
announce determinant formulas for the form factors of the operators $T_{13}(z)$ and $T_{31}(z)$.
To derive these formulas, we used a new approach, which requires a detailed description.
It will be given  in a separate publication.

The paper is organized  as follows. In section~\ref{S-N} we introduce the model under consideration.
In section~\ref{S-GL2} we recall the results for form factors in the models with $GL(2)$-symmetries.
In section~\ref{S-res} we present the main results of our paper. The methods of their derivation are
briefly described  in section~\ref{S-CFF}. In particular, we introduce there the notion of twisted
transfer matrix, which appears to be very effective for the calculation of form factors of the diagonal entries.
In section~\ref{S-proof} we present a proof of some determinant representations given in section~\ref{S-res}.
Section~\ref{S-Disc} is devoted to the discussions of some perspectives. Appendix collect several summation
identities needed for the proof of the determinant representations.

\section{Bethe vectors and form factors\label{S-N}}

In this section we describe the model under consideration, introduce necessary notations and
define the object of our study.

\subsection{Generalized $GL(3)$-invariant model\label{SS-mod}}

The models considered below are described by a $GL(3)$-invariant
$R$-matrix acting in the tensor product of two auxiliary spaces $V_1\otimes V_2$, where
$V_k\sim\mathbb{C}^3$, $k=1,2$:
 \be{R-mat}
 R(x,y)=\mathbf{I}+g(x,y)\mathbf{P},\qquad g(x,y)=\frac{c}{x-y}.
 \ee
In the above definition, $\mathbf{I}$ is the identity matrix in $V_1\otimes V_2$, $\mathbf{P}$ is the permutation matrix
that exchanges $V_1$ and $V_2$, and $c$ is  an arbitrary nonzero constant.

The monodromy matrix $T(w)$ satisfies the algebra
\be{RTT}
R_{12}(w_1,w_2)T_1(w_1)T_2(w_2)=T_2(w_2)T_1(w_1)R_{12}(w_1,w_2).
\ee
Equation \eqref{RTT} holds in the tensor product $V_1\otimes V_2\otimes\mathcal{H}$,
where $V_k\sim\mathbb{C}^3$, $k=1,2$, are the auxiliary linear spaces, and $\mathcal{H}$ is the Hilbert space of the Hamiltonian of the model under consideration. The  matrices $T_k(w)$ act non-trivially in
$V_k\otimes \mathcal{H}$.

The trace in the auxiliary space $V\sim\mathbb{C}^3$ of the monodromy matrix, $\tr T(w)$, is called the transfer matrix. It is a generating
functional of integrals of motion of the model. The eigenvectors of the transfer matrix are
called on-shell Bethe vectors (or simply on-shell vectors). They can be parameterized by sets of complex parameters
satisfying  Bethe equations (see section~\ref{SS-BV}).

\subsection{Notations\label{SS-N}}
We use the same notations and conventions as in the paper \cite{BelPRS13a}.
Besides the function $g(x,y)$ we also introduce a function $f(x,y)$
\be{univ-not}
 f(x,y)=1+g(x,y)=\frac{x-y+c}{x-y}.
\ee
Two other auxiliary functions  will be also used
\be{desand}
h(x,y)=\frac{f(x,y)}{g(x,y)}=\frac{x-y+c}{c},\qquad  t(x,y)=\frac{g(x,y)}{h(x,y)}=\frac{c^2}{(x-y)(x-y+c)}.
\ee
Due to the obvious property $g(-x,-y)=g(y,x)$ all the functions introduced above possess similar properties:
 \be{propert}
f(-x,-y)=f(y,x),\quad h(-x,-y)=h(y,x),\quad  t(-x,-y)=t(y,x).
 \ee

Before giving a description of the Bethe vectors we formulate a convention on the notations.
We denote sets of variables by bar: $\bar w$, $\bla$, $\bmu$ etc.
Individual elements
of the sets are denoted by subscripts: $w_j$, $\la_k$ etc. Notations $\bla_i$, $\bmu_i$
mean $\bla\setminus u_i$, $\bmu\setminus v_i$ etc.

In order to avoid too cumbersome formulas we use shorthand notations for products of
functions $g$, $f$, and $h$.  Namely, if these functions
depend on sets of variables, this means that one should take the product over the corresponding set.
For example,
 \be{SH-prod}
 h(z, \bar w)= \prod_{w_j\in\bar w} h(z, w_j);
 \quad  g(u_i, \bla_i)= \prod_{\substack{u_j\in\bar u\\u_j\ne u_i}} g(u_i, u_j);\quad
 f(\bla,\bmu)=\prod_{\la_j\in\bla}\prod_{\muu_k\in\bmu} f(\la_j,\muu_k).
 \ee
We will also use a special notation
$\Delta'_n(\bar x)$ and $\Delta_n(\bar x)$ for the products
\be{def-Del}
\Delta'_n(\bar x)
=\prod_{j<k}^n g(x_j,x_k),\qquad {\Delta}_n(\bar x)=\prod_{j>k}^n g(x_j,x_k).
\ee

\subsection{Bethe vectors\label{SS-BV}}

Now we pass to the description of Bethe vectors.
A generic Bethe vector is denoted by $\mathbb{B}^{a,b}(\bla;\bmu)$.
It is parameterized by two sets of
complex parameters $\bla=\la_1,\dots,\la_a$ and $\bmu=\muu_1,\dots,\muu_b$ with $a,b=0,1,\dots$.
Dual Bethe vectors are denoted by $\mathbb{C}^{a,b}(\bla;\bmu)$. They also depend on two sets of
complex parameters $\bla=\la_1,\dots,\la_{a}$ and $\bmu=\muu_1,\dots,\muu_{b}$. The state with
$\bla=\bmu=\emptyset$ is called a pseudovacuum vector $|0\rangle$. Similarly the dual state
with $\bla=\bmu=\emptyset$ is called a dual pseudovacuum vector $\langle0|$. These vectors
are annihilated by the operators $T_{ij}(w)$, where $i>j$ for  $|0\rangle$ and $i<j$ for $\langle0|$.
At the same time both vectors are eigenvectors for the diagonal entries of the monodromy matrix
 \be{Tjj}
 T_{ii}(w)|0\rangle=\as_i(w)|0\rangle, \qquad   \langle0|T_{ii}(w)=\as_i(w)\langle0|,
 \ee
where $\as_i(w)$ are some scalar functions. In the framework of the generalized model, $\as_i(w)$ remain free functional parameters. Actually, it is always possible to normalize
the monodromy matrix $T(w)\to \as_2^{-1}(w)T(w)$ so as to deal only with the ratios
 \be{ratios}
 r_1(w)=\frac{\as_1(w)}{\as_2(w)}, \qquad  r_3(w)=\frac{\as_3(w)}{\as_2(w)}.
 \ee

If the parameters $\bla$ and $\bmu$ of a Bethe vector\footnote{%
For simplicity here and below we do not distinguish between vectors and dual vectors.}
satisfy a special system of equations (Bethe equations), then
it becomes an eigenvector of the transfer matrix (on-shell Bethe vector). The system of Bethe equations can be written in the following form:
\be{AEigenS-1}
\begin{aligned}
r_1(\la_i)&=\frac{f(\la_i,\bla_{i})}{f(\bla_{i},\la_{i})}f(\bmu,\la_{i}),\qquad i=1,\dots,a,\\
r_3(\muu_{j})&=\frac{f(\bmu_{j},\muu_{j})}{f(\muu_{j},\bmu_{j})}f(\muu_{j},\bla),\qquad j=1,\dots,b,
\end{aligned}
\ee
and we recall that $\bla_i=\bla\setminus\la_i$ and $\bmu_j=\bmu\setminus\muu_j$.

If $\bla$ and $\bmu$ satisfy the system \eqref{AEigenS-1}, then
\be{Left-act}
\tr T(w)\mathbb{B}^{a,b}(\bla;\bmu) = \tau(w|\bla,\bmu)\,\mathbb{B}^{a,b}(\bla;\bmu),\qquad
\mathbb{C}^{a,b}(\bla;\bmu)\tr T(w) = \tau(w|\bla,\bmu)\,\mathbb{C}^{a,b}(\bla;\bmu),
\ee
where
\be{tau-def}
\tau(w)\equiv\tau(w|\bla,\bmu)=r_1(w)f(\bla,w)+f(w,\bla)f(\bmu,w)+r_3(w)f(w,\bmu).
\ee
{\sl Remark}. \label{Rem-1} Observe that the system of Bethe equations \eqref{AEigenS-1} is equivalent to the
statement that the  function $\tau(w|\bla,\bmu)$ \eqref{tau-def} has no poles in the points
$w=u_i$ and $w=v_j$.

Form factors of the monodromy matrix entries are defined as
 \be{SP-deFF-gen}
 \mathcal{F}_{a,b}^{(i,j)}(z)\equiv\mathcal{F}_{a,b}^{(i,j)}(z|\blac,\bmuc;\blab,\bmub)=
 \mathbb{C}^{a',b'}(\blac;\bmuc)T_{ij}(z)\mathbb{B}^{a,b}(\blab;\bmub),
 \ee
where both $\mathbb{C}^{a',b'}(\blac;\bmuc)$ and $\mathbb{B}^{a,b}(\blab;\bmub)$ are on-shell
Bethe vectors, and
\be{apabpb}
\begin{array}{l}
a'=a+\delta_{i1}-\delta_{j1},\\
b'=b+\delta_{j3}-\delta_{i3}.
\end{array}
\ee
We use here superscripts $B$ and $C$ in order to distinguish the sets of parameters
entering these two vectors. In other words, unless explicitly
specified, the variables $\{\blab; \bmub\}$ in $\mathbb{B}^{a,b}$ and
$\{\blac; \bmuc\}$ in $\mathbb{C}^{a,b}$ are not supposed to be  related.
The parameter $z$ is an arbitrary complex  number. Acting with the operator $T_{ij}(z)$ on $\mathbb{B}^{a,b}(\blab;\bmub)$
via formulas obtained in \cite{BelPRS12c} we reduce the form factor to a linear combination of
scalar products, in which $\mathbb{C}^{a',b'}(\blac;\bmuc)$ is an on-shell vector.

\subsection{Relations between form factors\label{SS-morph}}

Obviously, there exist nine form factors of $T_{ij}(z)$ in the models with $GL(3)$-invariant
$R$-matrix. However, not all of them are independent. In particular, due to the invariance of the
$R$-matrix under transposition with respect to both spaces, the mapping\footnote{For simplicity we denoted mappings
\eqref{def-psi}, \eqref{psiBV} and \eqref{psiFF} acting on the operators, vectors and form factors by the same letter $\psi$. The same is applied to the mappings \eqref{def-phi}, \eqref{phiBV} and \eqref{phiFF}.}
\be{def-psi}
\psi\,:
T_{ij}(u) \quad\mapsto\quad T_{ji}(u)
\ee
defines an antimorphism of the algebra \eqref{RTT}. Acting on the Bethe vectors this antimorphism maps them
into the dual ones and vice versa
\be{psiBV}
\psi \big(\mathbb{B}^{a,b}(\bar u;\bar v)\big) = \mathbb{C}^{a,b}(\bar u;\bar v),
\qquad \psi \big(\mathbb{C}^{a,b}(\bar u;\bar v)\big) = \mathbb{B}^{a,b}(\bar u;\bar v).
\ee
Therefore we have
\be{psiFF}
\psi \big(\mathcal{F}_{a,b}^{(i,j)}(z|\blac,\bmuc;\blab,\bmub)\big) =
\mathcal{F}_{a',b'}^{(j,i)}(z|\blab,\bmub;\blac,\bmuc),
\ee
where $a'$ and $b'$ are defined in \eqref{apabpb}.  Hence, the
form factor $\mathcal{F}_{a,b}^{(i,j)}(z)$ can be obtained from  $\mathcal{F}_{a,b}^{(j,i)}(z)$
by means of the replacements $\{\blac,\bmuc\}\leftrightarrow\{\blab,\bmub\}$ and $\{a,b\}\leftrightarrow\{a',b'\}$.

One more relationship
between form factors arises due to the mapping $\varphi$:
\be{def-phi}
\varphi\,:
T_{ij}(u) \quad\mapsto\quad T_{4-j,4-i}(-u),
\ee
that defines an isomorphism of the algebra \eqref{RTT} \cite{BelPRS12c}.
This isomorphism implies the following transform of Bethe vectors:
\be{phiBV}
\varphi \big(\mathbb{B}^{a,b}(\bar u;\bar v)\big) = \mathbb{B}^{b,a}(-\bar v;-\bar u),
\qquad \varphi \big(\mathbb{C}^{a,b}(\bar u;\bar v)\big) = \mathbb{C}^{b,a}(-\bar v;-\bar u).
\ee
Since the mapping $\varphi$ connects the operators $T_ {11}$ and $T_ {33}$, it also leads to the replacement of functions
$r_1 \leftrightarrow r_3$.  Thus,
\be{phiFF}
\varphi \big(\mathcal{F}_{a,b}^{(i,j)}(z|\blac,\bmuc;\blab,\bmub)\big) =
\mathcal{F}_{b,a}^{(4-j,4-i)}(-z|-\bmuc,-\blac;-\bmub,-\blab)\Bigr|_{r_1\leftrightarrow r_3}.
\ee
Altogether we are left with (at most) four independent form factors, for example, the form factors of the operators $T_{11}(z)$, $T_{22}(z)$, $T_{12}(z)$ and  $T_{13}(z)$.

\section{Form factors in $GL(2)$-based models\label{S-GL2}}

Before giving the main results of this paper we recall  the determinant representations for form factors
obtained previously in the integrable models with $GL(2)$-invariant $R$-matrix \cite{KojKS97,KitMT99}. Actually these
results can be treated as a particular cases of form factors in the models with $GL(3)$-invariant $R$-matrix,
which correspond to special Bethe vectors with $a=0$ or $b=0$.
Below we set for definiteness $b=0$. Let
\be{BV-gl2}
  \mathbb{C}^{a}(\bla) =\mathbb{C}^{a,0}(\bla;\emptyset),
  \qquad \mathbb{B}^{a}(\bla) =\mathbb{B}^{a,0}(\bla;\emptyset).
  \ee
The Bethe vectors \eqref{BV-gl2} become on-shell, if the parameters $\bla$ satisfy the system
of Bethe equations
\be{BE-gl2}
r_1(\la_i)=\frac{f(\la_i,\bla_{i})}{f(\bla_{i},\la_{i})}=(-1)^{a-1}\frac{h(\la_i,\bla)}{h(\bla,\la_i)},
\qquad i=1,\dots,a.
\ee
Then
\be{eigenval}
\bigl(T_{11}(w)+T_{22}(w)\bigr)\mathbb{B}^{a}(\bla) = \tau_2(w|\bla)\,\mathbb{B}^{a}(\bla),\qquad
\mathbb{C}^{a}(\bla)\bigl(T_{11}(w)+T_{22}(w)\bigr) = \tau_2(w|\bla)\,\mathbb{C}^{a}(\bla),
\ee
where
\be{tau2}
\tau_2(w)\equiv\tau_2(w|\bla)=r_1(w)f(\bla,w)+f(w,\bla).
\ee

The form factors of the monodromy matrix entries in the $GL(2)$-based models are defined as
 \be{def-FFgl2}
 \mathcal{F}_{a}^{(i,j)}(z)\equiv\mathcal{F}_{a}^{(i,j)}(z|\blac;\blab)=
 \mathbb{C}^{a'}(\blac)T_{ij}(z)\mathbb{B}^{a}(\blab),
 \ee
where both vectors are on-shell. For conciseness, we have used the notation $a'=a+j-i$.

All the representations for the form factors of the operators $T_{ij}(z)$, $i,j=1,2$, are based
on the determinant formula for the scalar product of on-shell Bethe vector and generic Bethe vector
\cite{Sla89}. This formula  immediately implies such the representations for $\mathcal{F}_{a}^{(1,2)}(z)$
and $\mathcal{F}_{a}^{(2,1)}(z)$. Namely, let $\bar x=\{\blab,z\}$. Then
\be{FF12-gl2}
\mathcal{F}_{a}^{(1,2)}(z)=\Delta'_{a'}(\blac)\Delta_{a'}(\bar x)\;
\det_{a'}\Ngltwo_{jk},
\ee
where
\be{N-gl2}
\Ngltwo_{jk}=\frac{c}{g(x_k,\blac)}
\frac{\partial\tau_2(x_k|\blac)}{\partial\lac_j}.
\ee
The result for  $\mathcal{F}_{a}^{(2,1)}(z)$ can be obtained from \eqref{FF12-gl2}, \eqref{N-gl2} via the
replacements $\blac\leftrightarrow\blab$ and $a'\leftrightarrow a$:
\be{FF21-gl2}
\mathcal{F}_{a}^{(2,1)}(z)=\Delta'_{a}(\blab)\Delta_{a}(\bar y)\;
\det_{a}\left(\frac{c}{g(y_k,\blab)}
\frac{\partial\tau_2(y_k|\blab)}{\partial\lab_j}\right),
\ee
where $\bar y=\{\blac,z\}$.


There exist several equivalent formulas for form factors of the diagonal entries $T_{ss}(z)$, $s=1,2$.
Here we give representations in the form of determinants of matrices of the size $(a+1)\times(a+1)$. We have
\be{FFss-gl2}
\mathcal{F}_{a}^{(s,s)}(z)=\Delta'_{a}(\blac)\Delta_{a+1}(\bar x)\;
\det_{a+1}\Ngltwo^{(s)}_{jk}, \qquad s=1,2,
\ee
where $\bar x=\{\blab,z\}$. 
The entries $\Ngltwo^{(s)}_{jk}$ of the matrices $\Ngltwo^{(s)}$ in the first $a$ rows ($j=1,\dots,a$) are given by \eqref{N-gl2}. Pay attention, however, that the cardinality of the set $\blac$ in
\eqref{N-gl2} is equal to $a+1$, while we have $\#\blac=a$ for the form factors $\mathcal{F}_{a}^{(s,s)}(z)$. One can say
that in both cases $\#\blac =a'$. 
%
In the last row one has
\be{Ns-gl2}
\begin{array}{l}
\Ngltwo^{(1)}_{a+1,k}=(-1)^ar_1(x_k)h(\blab,x_k),\\
\Ngltwo^{(2)}_{a+1,k}=h(x_k,\blab).
\end{array}
\ee

{\sl Remark}. Observe that due to Bethe equations \eqref{BE-gl2} we have $\Ngltwo^{(1)}_{a+1,k}
+\Ngltwo^{(2)}_{a+1,k}=0$ for $k=1,\dots,a$ (that is, if $x_k\in\blab$). Therefore the form factor
of the transfer matrix $T_{11}(z)+T_{22}(z)$ reduces to the eigenvalue $\tau_2(z|\blab)$ multiplied by
the scalar product of the vectors $\mathbb{C}^{a}(\blac)$ and $\mathbb{B}^{a}(\blab)$.
This result, of course, immediately follows from the definition of on-shell Bethe vectors.

Making the replacement $\blac\leftrightarrow\blab$ in \eqref{FFss-gl2}--\eqref{Ns-gl2} we obtain alternative
determinant representations for form factors of the operators $T_{ss}(z)$. In spite of these two types of
representations look very different, one can prove their equivalence (see e.g. \cite{KitKMST05b}).

Thus, we see that in the $GL(2)$-based models the form factors of the monodromy matrix entries are
proportional to the Jacobians of the eigenvalue $\tau_2(w)$ on the left or right Bethe vector (up to
possible modification of one row).

\section{Main results\label{S-res}}

The results given in  section~\ref{S-GL2} suggest their possible generalization to the models with
$GL(3)$-invariant $R$-matrix. Indeed, it seems quite reasonable to expect that form factors
of the monodromy matrix entries in such models also are proportional to the Jacobians of the transfer matrix eigenvalue.
However, this conjecture confirms only partly. In this section we show that the form factors of the
operators $T_{ij}(z)$ in the $GL(3)$-based models have more sophisticated determinant representations.

\subsection{Form factors of off-diagonal elements\label{S-off-d}}

The determinant representations for form factors of the operators $T_{ij}(z)$ with $|i-j|=1$ have the most simple structure.
They were calculated in \cite{PakRS14b}. We start our exposition with the form factor $ \mathcal{F}_{a,b}^{(1,2)}(z)$:
 \be{SP-deFF}
 \mathcal{F}_{a,b}^{(1,2)}(z)\equiv\mathcal{F}_{a,b}^{(1,2)}(z|\blac,\bmuc;\blab,\bmub)=
 \mathbb{C}^{a',b'}(\blac;\bmuc)T_{12}(z)\mathbb{B}^{a,b}(\blab;\bmub),
 \ee
where both $\mathbb{C}^{a',b'}(\blac;\bmuc)$ and $\mathbb{B}^{a,b}(\blab;\bmub)$ are on-shell
Bethe vectors.  As in the $GL(2)$ case, we used $a'$ and $b'$ notation, whose definition depends on the form factor we are considering. For the $\mathcal{F}_{a,b}^{(1,2)}(z)$ form factor, we have $a'=a+1$, $b'=b$.

In order to describe the determinant representation for this form factor we  introduce a
set of variables $\bar x=\{x_1,\dots,x_{a'+b}\}$ as the union of three sets $\bar x=\{\blab,\bmuc,z\}$,
and define a scalar function $\Hot_{a',b}$ as
\be{Hab}
\Hot_{a',b}=\frac{h(\bar x,\blab)h(\bmuc,\bar x)}{h(\bmuc,\blab)}\;\Delta'_{a'}(\blac)
\Delta'_{b}(\bmub)\Delta_{a+b+1}(\bar x).
\ee

\begin{prop}\label{PFF-12}(\cite{PakRS14b})
The form factor $\mathcal{F}_{a,b}^{(1,2)}(z)$ admits the following determinant representation:
\be{FF-ans-m}
\mathcal{F}_{a,b}^{(1,2)}(z)=\Hot_{a',b}\;
\det_{a'+b}\Not \,,
\ee
where  $(a'+b)\times(a'+b)$ matrix
$\Not$ has the  following entries

\begin{align}\label{Lu-mt}
\Not_{j,k}&=\frac{c}{f(x_k,\blab)f(\bmuc,x_k)}\frac{g(x_k,\blab)}{g(x_k,\blac)}
\frac{\partial\tau(x_k|\blac,\bmuc)}{\partial\lac_j},&\qquad j=1,\dots,a',\\
\label{Lv-mt}
\Not_{a'+j,k}&=\frac{-c}{f(x_k,\blab)f(\bmuc,x_k)}\frac{g(\bmuc,x_k)}{g(\bmub,x_k)}
\frac{\partial\tau(x_k|\blab,\bmub)}{\partial\mub_j},&\qquad j=1,\dots,b.
\end{align}
\end{prop}

We see that this representation involves two eigenvalues of the transfer matrix. Namely, the elements in the first
$a+1$ rows of the matrix $\Not$ are proportional to the derivatives of the eigenvalue $\tau(x_k|\blac,\bmuc)$
on the left vector, while the elements in the last
$b$ rows of the matrix $\Not$ are proportional to the derivatives of the eigenvalue $\tau(x_k|\blab,\bmub)$
on the right vector. Thus, as we have mentioned in the beginning of the section, this determinant representation
is not a straightforward generalization of the formula \eqref{FF12-gl2}. Nevertheless, one can easily see that
at $b=0$ the equation \eqref{FF-ans-m} reproduces the result \eqref{FF12-gl2}.

Determinant representations for other form factors $ \mathcal{F}_{a,b}^{(i,j)}(z)$ with $|i-j|=1$ can be
derived from \eqref{FF-ans-m} by  the mappings \eqref{psiFF}, \eqref{phiFF}. First, we give  the explicit
formulas for the form factor of the operator $T_{23}$
 \be{3SP-deFF}
 \mathcal{F}_{a,b}^{(2,3)}(z)\equiv\mathcal{F}_{a,b}^{(2,3)}(z|\blac,\bmuc;\blab,\bmub)=
 \mathbb{C}^{a',b'}(\blac;\bmuc)T_{23}(z)\mathbb{B}^{a,b}(\blab;\bmub),
 \ee
where  now for the $\mathcal{F}_{a,b}^{(2,3)}(z)$ form factor, we have $a'=a$ and $b'=b+1$.

We introduce  a set of variables $\bar y=\{y_1,\dots,y_{a+b'}\}$ as the union of three sets
$\bar y=\{\blac,\bmub,z\}$ and a function
\be{3Hab}
\Htt_{a,b'}=\frac{h(\bar y,\blac)h(\bmub,\bar y)}{h(\bmub,\blac)}\;\Delta'_{a}(\blab)
\Delta'_{b'}(\bmuc)\Delta_{a+b+1}(\bar y).
\ee

\begin{prop}\label{PFF-23}(\cite{PakRS14b})
The form factor $\mathcal{F}_{a,b}^{(2,3)}(z)$ admits the following determinant representation:
\be{3FF-ans-m}
\mathcal{F}_{a,b}^{(2,3)}(z)=\Htt_{a,b'}\;
\det_{a+b'}\Ntt\,,
\ee
where  $(a+b')\times(a+b')$ matrix
$\Ntt$ has the  following entries
\begin{align}\label{Lu-mt-m}
\Ntt_{j,k}&=\frac{c}{f(y_k,\blac)f(\bmub,y_k)}\frac{g(y_k,\blac)}{g(y_k,\blab)}
\frac{\partial\tau(y_k|\blab,\bmub)}{\partial\lab_j},&\qquad j=1,\dots,a,\\
\label{Lv-mt-m}
\Ntt_{a+j,k}&=\frac{-c}{f(y_k,\blac)f(\bmub,y_k)}\frac{g(\bmub,y_k)}{g(\bmuc,y_k)}
\frac{\partial\tau(y_k|\blac,\bmuc)}{\partial\muc_j},&\qquad j=1,\dots,b'.
\end{align}
\end{prop}

Using \eqref{propert} it is easy to check that the representation for $\mathcal{F}_{a,b}^{(2,3)}(z)$
can be obtained from the one for $\mathcal{F}_{a,b}^{(1,2)}(z)$ via the following replacements
\be{replac-1}
\blac\leftrightarrow-\bmuc,\quad \blab\leftrightarrow-\bmub,\quad r_1\leftrightarrow r_3,\quad
a\leftrightarrow b,
\ee
as it is prescribed by the isomorphism \eqref{phiFF}.

At the same time,  one can observe that the formulas for  these two form factors
are also related by the replacements
\be{replac-2}
\{\blac,\bmuc\}\leftrightarrow\{\blab,\bmub\},\quad \{a,b\}\leftrightarrow \{a',b'\}.
\ee
Be careful however that in doing these transformations, the definition of $a'$ and $b'$ changes when going from $\mathcal{F}_{a,b}^{(1,2)}(z)$ to $\mathcal{F}_{a,b}^{(2,3)}(z)$ (and vice-versa).

Applying mapping  \eqref{psiFF} to  representations \eqref{FF-ans-m}, \eqref{3FF-ans-m}
we arrive at the following

\begin{prop}\label{PFF-32-21}(\cite{PakRS14b})
The form factor $\mathcal{F}_{a,b}^{(3,2)}(z)$ admits the following determinant representation:
\be{32FF-ans-m}
\mathcal{F}_{a,b}^{(3,2)}(z)=\Hot_{a',b}\;
\det_{a'+b}\Not\,,
\ee
where $\Hot_{a',b}$ and $\Not$ are given by \eqref{Hab} and \eqref{FF-ans-m} respectively.

The form factor $\mathcal{F}_{a,b}^{(2,1)}(z)$ admits the following determinant representation:
\be{21FF-ans-m}
\mathcal{F}_{a,b}^{(2,1)}(z)=\Htt_{a,b'}\;
\det_{a+b'}\Ntt\,,
\ee
where $\Htt_{a,b'}$ and $\Ntt$ are given by \eqref{3Hab} and \eqref{3FF-ans-m} respectively.
\end{prop}

{\sl Remark}. We would like to stress  again that although the representations \eqref{32FF-ans-m} and \eqref{21FF-ans-m}
formally coincide with \eqref{FF-ans-m} and \eqref{3FF-ans-m}, the values of $a'$ and $b'$ in these formulas are different.
Indeed, one has $a'=a+1$ and $b'=b$ in \eqref{FF-ans-m}, while $a'=a$ and $b'=b-1$ in \eqref{32FF-ans-m}. Similarly
$a'=a$ and $b'=b+1$ in \eqref{3FF-ans-m}, while $a'=a-1$ and $b'=b$ in \eqref{21FF-ans-m}. Therefore, in particular,
the matrices $\Not$ and $\Ntt$ in \eqref{FF-ans-m} and \eqref{3FF-ans-m} have a size
$(a+b+1)\times(a+b+1)$, while in the equations \eqref{32FF-ans-m} and \eqref{21FF-ans-m} the same
matrices have a size $(a+b)\times(a+b)$.

\subsection{Form factors of diagonal elements\label{S-d}}

The form factors of diagonal entries of the monodromy matrix
 \be{SP-deFFss}
 \mathcal{F}_{a,b}^{(s,s)}(z)\equiv\mathcal{F}_{a,b}^{(s,s)}(z|\blac,\bmuc;\blab,\bmub)=
 \mathbb{C}^{a,b}(\blac;\bmuc)T_{ss}(z)\mathbb{B}^{a,b}(\blab;\bmub),
 \ee
were calculated in \cite{BelPRS13a}. Here we give different representations for them. In a sense they
are analogous to the determinant formulas for form factors in the $GL(2)$-based models (see section~\ref{S-GL2}).
Namely, they are  based on the determinant of the matrix $\Not$ \eqref{Lu-mt}, \eqref{Lv-mt}, but one
row of this matrix should be modified.

As before we combine the sets $\blab$ and $\bmuc$ and the parameter $z$ into the set  $\bar x=\{\blab,\bmuc,z\}$.
We also introduce three $(a+b+1)$-component vectors $Y^{(s)}$, $s=1,2,3$, as
\be{Y1}
\begin{aligned}
Y^{(s)}_k&=\delta_{s2}-\delta_{s1}+\frac{\lab_k}{c}(\delta_{s1}-\delta_{s3})
\left(\frac{f(\bmub,\lab_k)}{f(\bmuc,\lab_k)}-1\right),\qquad k=1,\dots,a;\num
Y^{(s)}_{a+k}&=\delta_{s2}-\delta_{s3}+\frac{\muc_k+c}{c}(\delta_{s1}-\delta_{s3})
\left(\frac{f(\muc_k,\blac)}{f(\muc_k,\blab)}-1\right),\qquad k=1,\dots,b.
\end{aligned}
\ee
In these formulas  $\delta_{sk}$ are Kronecker  deltas.
The values of $Y^{(s)}_{a+b+1}$ are crucial only in the case when $ \mathbb{C}^{a,b}(\blac;\bmuc)=
\left(\mathbb{B}^{a,b}(\blab;\bmub)\right)^\dagger$,
that is $\blab=\blac=\bla$ and $\bmub=\bmuc=\bmu$. We define them as
\be{Y2}
Y^{(1)}_{a+b+1}=\frac{r_1(z)f(\bla,z)}{f(\bmu,z)f(z,\bla)},\qquad
Y^{(2)}_{a+b+1}=1,\qquad
Y^{(3)}_{a+b+1}=\frac{r_3(z)f(z,\bmu)}{f(\bmu,z)f(z,\bla)}.
\ee
One can set here
$\bmu=\bmuc$ or $\bmu=\bmub$, as well as $\bla=\blac$ or $\bla=\blab$.

\begin{prop}\label{PFF-ss}
Define an $(a+b+1)\times(a+b+1)$ matrix $\Not^{(s)}$ as follows
\be{L-def}
\begin{aligned}
&\Not_{j,k}^{(s)}=\Not_{j,k},\qquad &j=1,\dots,a+b;\\
&\Not_{a+b+1,k}^{(s)}=Y_k^{(s)}.
\end{aligned}
\ee
Here the matrix $\Not$ is given by \eqref{Lu-mt}, \eqref{Lv-mt}. Then
 \be{SP-res}
 \mathcal{F}_{a,b}^{(s,s)}(z)=(-1)^b \mathcal{H}_{a',b}\cdot\det_{a+b+1}\Not^{(s)},
 \ee
where $\mathcal{H}_{a',b}$ is given by \eqref{Hab}.
\end{prop}

{\sl Remark}. One should remember that in the case of the
form factors  $\mathcal{F}_{a,b}^{(s,s)}(z)$ one has $a'=a$, while $a'=a+1$ in the case of the form factor $\mathcal{F}_{a,b}^{(1,2)}(z)$. Therefore the function $\mathcal{H}_{a',b}$ in \eqref{SP-res} is given by
\eqref{Hab}, where one should set $a'=a$. The same remark concerns the entries of the matrix $\Not^{(s)}$.

We prove this proposition in section~\ref{S-proof}, reducing the representation \eqref{SP-res} to the formulas obtained
in \cite{BelPRS13a}. However before doing this we would like to mention that similarly to the $GL(2)$-case representation
\eqref{SP-res} implies several alternative determinant formulas for the form factors of the diagonal entries of the monodromy matrix. They can be obtained from \eqref{SP-res} via the morphisms \eqref{psiFF} and \eqref{phiFF}.

It is also worth mentioning that
\be{sum-Y-s}
\begin{aligned}
\sum_{s=1}^3Y^{(s)}_{k}&=0,\qquad k=1,\dots,a+b,\\
\sum_{s=1}^3Y^{(s)}_{a+b+1}&=\frac{\tau(z|\bla,\bmu)}{f(z,\bla)f(\bmu,z)}.
\end{aligned}
\ee
Therefore the form factor of the transfer matrix reduces to its eigenvalue $\tau(z|\bla,\bmu)$ multiplied by
the minor of the matrix $\Not^{(s)}$ built on the first $(a+b)$ rows and columns. This minor vanishes, if
$ \mathbb{C}^{a,b}(\blac;\bmuc)\ne\left(\mathbb{B}^{a,b}(\blab;\bmub)\right)^\dagger$ (see \cite{BelPRS12b}
and section~\ref{SS-Pds}), and thus, the form factor of the transfer matrix between different states is equal to
zero, as it should be. Otherwise, if $ \mathbb{C}^{a,b}(\blac;\bmuc)=\left(\mathbb{B}^{a,b}(\blab;\bmub)\right)^\dagger$,
then  the form factor of the transfer matrix is equal to the  eigenvalue $\tau(z|\bla,\bmu)$ multiplied by
square of the norm of Bethe vector (see section~\ref{SS-PSs}).

\subsection{Form factor of $T_{13}(z)$\label{S-FFT13}}

The form factors of  the matrix element $T_{13}(z)$ is defined as
 \be{13SP-deFFss}
 \mathcal{F}_{a,b}^{(1,3)}(z)\equiv\mathcal{F}_{a,b}^{(1,3)}(z|\blac,\bmuc;\blab,\bmub)=
 \mathbb{C}^{a',b'}(\blac;\bmuc)T_{13}(z)\mathbb{B}^{a,b}(\blab;\bmub),
 \ee
where $a'=a+1$ and $b'=b+1$.  As already mentioned, the calculation of this form factor relies on a new method,
that  will be presented elsewhere. However, to have here a complete overview of the form factors of the $GL(3)$ case, we preview the result. The  determinant representation of the $T_{13}(z)$ form factor
 is similar to the ones for the form factors of the diagonal
entries $T_{ss}(z)$. We again combine the sets $\blab$ and $\bmuc$ and the parameter $z$ into the set  $\bar x=\{\blab,\bmuc,z\}$. However now this set contains $a'+b'$ (that is, $a+b+2$) elements.
We also introduce  $(a'+b')$-component vector $Y^{(1,3)}$ as
\be{Y13}
Y^{(1,3)}_k=(-1)^{b'}\frac{r_3(x_k)h(x_k,\bmub)}{f(x_k,\blab)h(\bmuc,x_k)}
+\frac{h(\bmub,x_k)}{h(\bmuc,x_k)}.
\ee

\begin{prop}\label{PFF-13}
Define an $(a'+b')\times(a'+b')$ matrix $\Not^{(1,3)}$ as follows
\be{13L-def}
\begin{aligned}
&\Not_{j,k}^{(1,3)}=\Not_{j,k},\qquad &j=1,\dots,a'+b;\\
&\Not_{a'+b',k}^{(1,3)}=Y_k^{(1,3)}.
\end{aligned}
\ee
Here the matrix $\Not$ is given by \eqref{Lu-mt}, \eqref{Lv-mt}. Then
 \be{13SP-res}
 \mathcal{F}_{a,b}^{(1,3)}(z)=(-1)^{b'} \mathcal{H}_{a',b}\cdot\det_{a'+b+1}\Not^{(1,3)},
 \ee
where $\mathcal{H}_{a',b}$ is given by \eqref{Hab}.
\end{prop}

Note that one can obtain an alternative determinant representation for the form factor $\mathcal{F}_{a,b}^{(1,3)}(z)$
applying the mapping \eqref{phiFF} to the result \eqref{13SP-res}. In its turn, the application of the antimorphism \eqref{psiFF} to  \eqref{13SP-res} leads us to a determinant representation for the form factor
$\mathcal{F}_{a,b}^{(3,1)}(z)$.

\section{Calculation of form factors\label{S-CFF}}

As we have mentioned already, the determinant representation for the scalar product of on-shell Bethe vector
and generic Bethe vector plays a key role in calculating form factors in $GL(2)$-based models. In the
case of the $GL(3)$ group, an analog of such determinant representation is not known. Therefore calculating the form factor
becomes much more involved. The reader can find the details of these calculations in papers \cite{BelPRS12b,BelPRS13a,PakRS14b}. Here we give only a general description of the method that we have used in the papers above.

The study of form factors is based on an explicit representation for the scalar products
of Bethe vectors obtained in \cite{Res86,Whe12,BelPRS12a}.  The scalar product is defined as
 \be{SP-def}
S_{a,b}\equiv S_{a,b}(\blac,\bmuc;\blab,\bmub)=
 \mathbb{C}^{a,b}(\blac;\bmuc)\mathbb{B}^{a,b}(\blab;\bmub).
 \ee
Here the Bethe parameters $\{\blac\,,\,\bmuc\}$ and  $\{\blab\,,\,\bmub\}$ are supposed to be generic
complex numbers. The representation obtained in \cite{Res86} describes the scalar
product as a sum over partitions of Bethe parameters into subsets (so called {\it sum formula}).
 Generically this representation is not reducible to a more compact form. However, when calculating
the form factors, we deal with very particular  scalar products, where most of the parameters satisfy the Bethe equations \eqref{AEigenS-1}. In such cases, one can reduce this sum over partitions to a single determinant.

Consider, for example, the form
factor of the operator $T_{12}(z)$. The action of $T_{12}(z)$ onto $\mathbb{B}^{a,b}(\blab;\bmub)$
is (see \cite{BelPRS12c})
 \begin{align}
 T_{12}(z)\mathbb{B}^{a,b}(\blab;\bmub)&= f(\bmub,z)\, \mathbb{B}^{a+1,b}(\{\blab,z\};\bmub)\\
&+ \sum_{i=1}^b g(z,\mub_i) f(\bmub_i,\mub_i)\, \mathbb{B}^{a+1,b}(\{\blab,z\};\{\bmub_i,z\}).
 \label{act12}\end{align}
Thus, the form factor of  $T_{12}(z)$ is equal to
 \begin{align}
\mathcal{F}_{a,b}^{(1,2)}(z)&= f(\bmub,z)\, S_{a+1,b}(\blac,\bmuc;\{\blab,z\},\bmub)\\
&+ \sum_{i=1}^b g(z,\mub_i) f(\bmub_i,\mub_i)\, S_{a+1,b}(\blac,\bmuc;\{\blab,z\},\{\bmub_i,z\}),
 \label{FF-SP}
 \end{align}
and we have reduced the original problem to the calculation of the scalar products, where  only $z$ is an arbitrary
complex number, while other variables satisfy Bethe equations \eqref{AEigenS-1}.

Formally, other form factors can be calculated in a similar manner.
It was proved in \cite{BelPRS12c} that the action of the monodromy matrix entries on Bethe vectors
reduces to a linear combination of the last ones. Thus, the form factors of $T_{ij}(z)$ always can be
expressed in terms of linear combination of scalar products.
However, every specific case has its own peculiarities. In particular, as we have explained above,
there is no need to perform a special consideration of form factors $\mathcal{F}_{a,b}^{(i,j)}(z)$
with $|i-j|=1$, as all of them can be obtained from $\mathcal{F}_{a,b}^{(1,2)}(z)$ via the mappings
\eqref{psiFF}, \eqref{phiFF}.

The form factors of the diagonal operators  $T_{ss}(z)$,
also  can be calculated  in the framework of the scheme described above. However the action
of $T_{ss}(z)$ onto $\mathbb{B}^{a,b}(\blab;\bmub)$ is much more involved than \eqref{act12}. In particular,
it contains a double sum over the Bethe parameters. This fact makes the straightforward calculation of
$\mathcal{F}_{a,b}^{(s,s)}(z)$  very complex from a technical viewpoint. Therefore, in the case of
form factors of the diagonal entries of the monodromy matrix, it is more convenient to apply a special
trick, based on the use of the {\it twisted transfer matrix}. We describe this method in the next subsection.

Finally, the calculation of form factors $\mathcal{F}_{a,b}^{(i,j)}(z)$
with $|i-j|=2$ also should be included into the general scheme. However, in this case we did not
succeed to perform the summation over partitions to a single determinant, because of
technical problems. It seems rather strange, because the action of the operator $T_{13}(z)$ on the Bethe vectors is the most simple
 \begin{equation}
 T_{13}(z)\mathbb{B}^{a,b}(\blab;\bmub)=  \mathbb{B}^{a+1,b+1}(\{\blab,z\};\{\bmub,z\}),
 \label{act13}\end{equation}
and therefore the form factor of  $T_{13}(z)$ is given by a single scalar product:
 \begin{equation}
\mathcal{F}_{a,b}^{(1,3)}(z)=  S_{a+1,b+1}(\blac,\bmuc;\{\blab,z\},\{\bmub,z\}).
 \label{FF-SP13}
 \end{equation}
Nevertheless,
in spite of this simplicity the method of calculation of the sums over partitions of Bethe parameters arising in
\eqref{FF-SP13} is not developed for today.
Therefore for the study the form factor $\mathcal{F}_{a,b}^{(i,j)}(z)$ with $|i-j|=2$ we use another approach, which
will be described in a separate publication.
Here we would like to mention only that the form factors
$\mathcal{F}_{a,b}^{(1,3)}(z)$ and $\mathcal{F}_{a,b}^{(3,1)}(z)$ are related by the
mapping \eqref{psiFF}.

\subsection{Twisted transfer matrix\label{SS-TTM}}

$GL(3)$-invariance of $R$-matrix \eqref{R-mat} means that $[\hat\kappa_1\hat\kappa_2,R_{12}]=0$
for arbitrary $\hat\kappa \in GL(3)$. It is easy to see \cite{Kor82,IzeK84,Res86,KitMST05} that due to this property a {\it twisted monodromy matrix} $\hat\kappa T(w)$ satisfies the algebra \eqref{RTT}. If the matrix $\hat\kappa T(w)$ possesses the
same pseudovacuum and dual pseudovacuum vectors as the original matrix $T(w)$, then one can apply all the tools of  the nested algebraic Bethe ansatz to the twisted monodromy matrix. In particular, one can find the spectrum of the
twisted transfer matrix $\tr \hat\kappa T(w)$. Its eigenvectors  are called twisted on-shell Bethe vectors (or simply twisted on-shell vectors).

Consider a matrix $\hat\kappa=\diag(\kappa_1,\kappa_2,\kappa_3)$, where $\kappa_i$ are
arbitrary complex numbers. Obviously, the corresponding twisted monodromy matrix has the
same pseudovacuum and dual pseudovacuum vectors. Actually, the multiplication of $T(w)$ by $\hat\kappa$
reduces to the replacement of the original eigenvalues $\lambda_i(w)$ \eqref{Tjj} by
$\kappa_i\lambda_i(w)$. Therefore, like the standard on-shell vectors, the twisted on-shell vectors can be parameterized by a set of complex parameters satisfying the twisted Bethe equations. The last ones have the form \eqref{AEigenS-1},
where one should replace $r_k(z)$ by $r_k(z)\;\kappa_k/\kappa_2$. Below we will need these equations in the
logarithmic form. Namely, let
\begin{align}\label{Phi-1}
\Phi_j&=\log r_1(\la_{j})-\log \left(\frac{f(\la_{j},\bla_{j})}{f(\bla_{j},\la_{j})}\right) -\log f(\bmu,\la_{j}),
\qquad j=1,\dots,a,\\
\Phi_{a+j}&=\log r_3(\muu_{j})-\log \left(\frac{f(\bmu_{j},\muu_{j})}{f(\muu_{j},\bmu_{j})}\right)-\log f(\muu_{j},\bla),
\qquad j=1,\dots,b.
\label{Phi-2}
\end{align}
Then the system of twisted Bethe equations has the form
\be{Log-TBE}
\begin{array}{ll}
\Phi_j=\log\kappa_2-\log\kappa_1+2\pi i \ell_j,\qquad & j=1,\dots,a,\\
\Phi_{a+j}=\log\kappa_2-\log\kappa_3+2\pi i m_j,\qquad & j=1,\dots,b,
\end{array}
\ee
where $\ell_j$ and $m_j$ are some integers. The Jacobian of \eqref{Phi-1} and \eqref{Phi-2} is closely related to the
norm of the on-shell Bethe vector and the average values of the operators $T_{ss}(z)$ \cite{BelPRS13a}.

Using the notion of the twisted transfer matrix one can calculate the form factors of the diagonal entries
of the monodromy matrix.
Consider the expectation value
\be{Qm}
Q_{\bar\kappa}(z)=\mathbb{C}_{\bar\kappa}^{a,b}(\blac;\bmuc) \bigl(\tr \hat\kappa T(z)-\tr T(z)\bigr)\mathbb{B}^{a,b}(\blab;\bmub),
\ee
where $\mathbb{C}_{\bar\kappa}^{a,b}(\blac;\bmuc)$ and $\mathbb{B}^{a,b}(\blab;\bmub)$ are twisted and standard on-shell
vectors respectively. Here and below we denote $\bar\kappa=\{\kappa_1,\kappa_2,\kappa_3\}$.  Obviously
\be{Qm-0}
Q_{\bar\kappa}(z)=\mathbb{C}_{\bar\kappa}^{a,b}(\blac;\bmuc)\sum_{j=1}^3(\kappa_j-1)T_{jj}(z) \mathbb{B}^{a,b}(\blab;\bmub),
\ee
and therefore
\be{Qm-00}
\frac{d Q_{\bar\kappa}(z)}{d\kappa_s}\Bigl.\Bigr|_{\bar\kappa=1}=
\mathbb{C}_{\bar\kappa}^{a,b}(\blac;\bmuc)\Bigl.\Bigr|_{\bar\kappa=1}T_{ss}(z) \mathbb{B}^{a,b}(\blab;\bmub).
\ee
Here $\bar\kappa=1$ means that $\kappa_i=1$ for $i=1,2,3$.  Observe that after setting  $\bar\kappa=1$
the  vector $\mathbb{C}_{\bar\kappa}^{a,b}(\blac;\bmuc)$ turns into the standard on-shell vector $\mathbb{C}^{a,b}(\blac;\bmuc)$.
Hence, we obtain the form factor of $T_{ss}(z)$ in the r.h.s. of \eqref{Qm-00}
\be{Qm-FF}
\frac{d Q_{\bar\kappa}(z)}{d\kappa_s}\Bigl.\Bigr|_{\bar\kappa=1}=
\mathcal{F}^{(s,s)}_{a,b}(z|\blac,\bmuc;\blab,\bmub).
\ee

On the other hand
\be{Qm-1}
Q_{\bar\kappa}(z)=\bigl(\tau_{\bar\kappa}(z|\blac;\bmuc)-\tau(z|\blab;\bmub)\bigr)
\;\mathbb{C}_{\bar\kappa}^{a,b}(\blac;\bmuc)\mathbb{B}^{a,b}(\blab;\bmub),
\ee
where $\tau(z|\blab;\bmub)$ is the eigenvalue of $\tr T(z)$ \eqref{tau-def}, while
$\tau_{\bar\kappa}(z|\blac;\bmuc)$  is the eigenvalues of the twisted transfer matrix  $\tr \hat\kappa T(z)$:
\be{tau-tw-def}
\tau_{\bar\kappa}(z)\equiv\tau_{\bar\kappa}(z|\bla,\bmu)=
\kappa_1r_1(z)f(\bla,z)+\kappa_2f(z,\bla)f(\bmu,z)+\kappa_3r_3(z)f(z,\bmu).
\ee
Thus, we obtain
\be{FF-twistSP}
\mathcal{F}^{(s,s)}_{a,b}(z)=
\frac{d}{d\kappa_s}
\Big[\bigl(\tau_{\bar\kappa}(z|\blac;\bmuc)-\tau(z|\blab;\bmub)\bigr)
\;\mathbb{C}_{\bar\kappa}^{a,b}(\blac;\bmuc)\mathbb{B}^{a,b}(\blab;\bmub)\Big]\Bigl.\Bigr|_{\bar\kappa=1},
\ee
and we see that the form factors $\mathcal{F}^{(s,s)}_{a,b}(z)$ can be calculated as $\kappa$-derivatives
of the scalar product between twisted on-shell and standard on-shell vectors.

\section{Proof of proposition~\ref{PFF-ss}\label{S-proof}}

In this section we prove proposition~\ref{PFF-ss}. More precisely, we show that the determinant
representations given by proposition~\ref{PFF-ss} are equivalent to the ones obtained in \cite{BelPRS13a}.

Dealing with the form factors of diagonal entries $T_{ss}(z)$ one should distinguish between two cases:
\begin{itemize}
\item
 $\mathbb{C}^{a,b}(\blac;\bmuc)\ne
\bigl(\mathbb{B}^{a,b}(\blab;\bmub)\bigr)^\dagger$;
\item
$\mathbb{C}^{a,b}(\blac;\bmuc)=
\bigl(\mathbb{B}^{a,b}(\blab;\bmub)\bigr)^\dagger$.
\end{itemize}

We consider these two cases separately.

\subsection{Proof for different states\label{SS-Pds}}

In this section $\mathbb{C}^{a,b}(\blac;\bmuc)\ne
\bigl(\mathbb{B}^{a,b}(\blab;\bmub)\bigr)^\dagger$. It means that
there exists at least one $w\in\{\blac,\bmuc\}$, such that $w\notin\{\blab,\bmub\}$.
Then
\be{Orthogon}
\mathbb{C}_{\bar\kappa}^{a,b}(\blac;\bmuc)\mathbb{B}^{a,b}(\blab;\bmub)\Bigl.\Bigr|_{\bar\kappa=1}=0,
\ee
as a product of two eigenstates corresponding to the  different eigenvalues of the
transfer matrix. Hence, the $\kappa$-derivative in \eqref{FF-twistSP} should be applied
only to this scalar product. We obtain
\be{FF-diff-st}
\mathcal{F}^{(s,s)}_{a,b}(z)=
\bigl(\tau(z|\blac;\bmuc)-\tau(z|\blab;\bmub)\bigr)\frac{d}{d\kappa_s}
\;\mathbb{C}_{\bar\kappa}^{a,b}(\blac;\bmuc)\mathbb{B}^{a,b}(\blab;\bmub)\Bigl.\Bigr|_{\bar\kappa=1}.
\ee

The $\kappa$-derivatives of the scalar product between twisted on-shell and standard on-shell vectors
were calculated in \cite{BelPRS13a}. Let us describe this result.

First of all we introduce an $(a+b)$-component vector $\Omega$ as
\be{def-Omega}
\begin{array}{l}
{\dis \Omega_j=\frac{g(\lac_j,\blac_j)}{g(\lac_j,\blab)}
,\qquad j=1,\dots,a,}\num
{\dis \Omega_{a+j}=\frac{g(\mub_j,\bmub_j)}{g(\mub_j,\bmuc)},\qquad j=1,\dots,b.}
\end{array}
\ee
It is easy to see that since $\mathbb{C}^{a,b}(\blac;\bmuc)\ne
\bigl(\mathbb{B}^{a,b}(\blab;\bmub)\bigr)^\dagger$, this vector has at least one non-zero component. Without loss of generality we assume that $\Omega_{a+b}\ne 0$. Then the result for the $\kappa$-derivative of the scalar product reads
\be{K-derSP}
\frac{d}{d\kappa_s}
\;\mathbb{C}_{\bar\kappa}^{a,b}(\blac;\bmuc)\mathbb{B}^{a,b}(\blab;\bmub)\Bigl.\Bigr|_{\bar\kappa=1}
=\Omega^{-1}_{a+b}\;H_{a,b}\;\Nmod_{a+b,a+b+1}.
\ee
Here
\be{hab-diff}
H_{a,b}=\frac{(-1)^b\mathcal{H}_{a,b}}{f(z,\blab)f(\bmuc,z)}=
\frac{h(\bar w,\blab)h(\bmuc,\bar w)}{h(\bmuc,\blab)}\;\Delta'_{a}(\blac)
\Delta'_{b}(\bmub)\Delta_{a+b}(\bar w),
\ee
where $\mathcal{H}_{a,b}$ is given by \eqref{Hab} and $\bar w=\{\blab,\bmuc\}$. The factor
$\Nmod_{a+b,a+b+1}$ in \eqref{K-derSP} is the cofactor to the element $\Not^{(s)}_{a+b,a+b+1}$
of the matrix $\Not^{(s)}$ \eqref{L-def}
\be{cofact}
\Nmod_{a+b,a+b+1}=-\det_{\substack{j\ne a+b\\k\ne a+b+1}}\Not^{(s)}_{j,k}.
\ee

Let us  reproduce this result starting from the determinant representation \eqref{SP-res}.
First of all we give the entries of the matrix $\Not$ more explicitly
\begin{align}\label{Lu-m}
\Not_{j,k}&=(-1)^{a'-1}t(\lac_j,x_k)\frac{r_1(x_k)h(\blac,x_k)}{f(\bmuc,x_k)h(x_k,\blab)}
+t(x_k,\lac_j)\frac{h(x_k,\blac)}{h(x_k,\blab)},&\quad j=1,\dots,a',\\
\label{Lv-m}
\Not_{a'+j,k}&=(-1)^{b-1}t(x_k,\mub_j)\frac{r_3(x_k)h(x_k,\bmub)}{f(x_k,\blab)h(\bmuc,x_k)}
+t(\mub_j,x_k)\frac{h(\bmub,x_k)}{h(\bmuc,x_k)},&\quad j=1,\dots,b.
\end{align}
Note that in the case under consideration $a'=a$ and $b'=b$. We use, however, the symbol $a'$
in \eqref{Lu-m}, \eqref{Lv-m}, because in this form the equations above are still valid for
form factor $\mathcal{F}^{(1,2)}_{a,b}(z)$, where $a'=a+1$.

Let
\be{sum-O}
S(x_k)=\sum_{j=1}^{a'}\Omega_j \Not_{j,k}
+\sum_{j=1}^b\Omega_{a'+j}\Not_{a'+j,k}.
\ee
Then using \eqref{tuz} one can easily find
\be{sum-O-res}
S(x_k)=
\frac{\tau(x_k|\blac,\bmuc)-\tau(x_k|\blab,\bmub)}{f(\bmuc,x_k)f(x_k,\blab)}.
\ee
It is straightforward to check that $S(\lab_k)=S(\muc_k)=0$ due to the Bethe equations. In fact,
one can see this without any calculations. Indeed, the Bethe
equations are equivalent to the statement that the function $\tau(x_k|\bla,\bmu)$ has no poles
in the points $x_k=u_j$ and $x_k=v_j$ (see Remark on the page~\pageref{Rem-1}). Then the factor $f^{-1}(\bmuc,x_k)f^{-1}(x_k,\blab)$
immediately yields the equalities $S(\lab_k)=S(\muc_k)=0$.

Now we multiply the first $(a+b-1)$ rows of the matrix $\Not^{(s)}$ by the factors $\Omega_j/\Omega_{a+b}$
and add them to the $(a+b)$-th row. Then we obtain a modified $(a+b)$-th row with the components
\be{L-mod}
\begin{aligned}
&\Not^{(s),\rm{mod}}_{a+b,k}=0,\qquad k=1,\dots,a+b,\num
&\Not^{(s),\rm{mod}}_{a+b,a+b+1}=\Omega^{-1}_{a+b}\frac{\tau(z|\blac,\bmuc)-\tau(z|\blab,\bmub)}{f(\bmuc,z)f(z,\blab)}.
\end{aligned}
\ee
The determinant $\det\Not^{(s)}$ reduces to the product of the element $\Not^{(s),\rm{mod}}_{a+b,a+b+1}$
by the corresponding cofactor, and we arrive at
\be{det-det}
\det_{a+b+1}\Not^{(s)}=\Omega^{-1}_{a+b}\frac{\tau(z|\blac,\bmuc)-\tau(z|\blab,\bmub)}{f(\bmuc,z)f(z,\blab)}
\Nmod_{a+b,a+b+1}.
\ee
We would like to draw the reader's attention that the matrix element $Y_{a+b+1}^{(s)}$ has disappeared from the game.
Substituting this result into \eqref{SP-res} we immediately reproduce \eqref{K-derSP}.

\subsection{Proof for the same states\label{SS-PSs}}

In this section $\mathbb{C}^{a,b}(\blac;\bmuc)=
\bigl(\mathbb{B}^{a,b}(\blab;\bmub)\bigr)^\dagger$ and we set
$\blac=\blab=\bla$ and $\bmuc=\bmub=\bmu$. In this case
\be{diff-tau}
\tau_{\bar\kappa}(z|\blac;\bmuc)-\tau(z|\blab;\bmub)=0,\qquad\text{at}\qquad
\bar\kappa=1; \quad \blac=\blab=\bla;\quad \bmuc=\bmub=\bmu,
\ee
hence, the $\kappa$-derivative in \eqref{FF-twistSP} should act only on the difference
of the eigenvalues $\tau_{\bar\kappa}$ and $\tau$. Then we find
\be{Qm-2}
\mathcal{F}^{(s,s)}(z|\bla,\bmu;\bla,\bmu)=
\|\mathbb{B}^{a,b}(\bla;\bmu)\|^2\;\frac{d \tau_{\bar\kappa}(z|\blac;\bmuc)}{d\kappa_s}\Bigl.\Bigr|_{\bar\kappa=1},
\ee
and  one should set $\blac=\bla$ and $\bmuc=\bmu$ after taking the derivative of $\tau_{\bar\kappa}(z|\blac;\bmuc)$
with respect to  $\kappa_s$. Below in this section we always assume that the condition $\bar\kappa=1$ automatically
yields $\blac=\blab=\bla$ and $\bmuc=\bmub=\bmu$.

The square of the norm of on-shell Bethe vector $\|\mathbb{B}^{a,b}(\bla;\bmu)\|^2$ was calculated in
\cite{Res86,BelPRS12b}. It is proportional to the minor of the matrix $\Not^{(s)}$ built on the
first $(a+b)$ rows and columns\footnote{Pay attention that this minor does not depend on $s$.}:
\be{Norm}
\|\mathbb{B}^{a,b}(\bla;\bmu)\|^2= H_{a,b}\det_{a+b}\Not,
 \ee
where $H_{a,b}$ is given by \eqref{hab-diff}  at $\blac=\blab=\bla$ and $\bmuc=\bmub=\bmu$.

Let us give explicitly the entries of the
matrix $\Not$ in the case $\blac=\blab=\bla$ and $\bmuc=\bmub=\bmu$ (see \cite{Res86,BelPRS12b}). For $j,k=1,\dots,a$ we have
 \begin{equation}\label{P11}
\Not_{j,k}=
 \delta_{jk}\left(-c\log'r_1(u_k)-\sum_{\ell=1}^a\frac{2c^2}{\la_{k\ell}^2-c^2}+\sum_{m=1}^b
 t(\muu_m,\la_k)\right)  +\frac{2c^2}{\la_{jk}^2-c^2},
 \end{equation}
where $u_{k\ell}=u_k-u_\ell$. The entries of the second diagonal block are
 \begin{equation}\label{P22}
\Not_{a+j,a+k}=\delta_{jk}\left(c\log'r_3(v_k)-\sum_{m=1}^b\frac{2c^2}{\muu_{km}^2-c^2}+\sum_{\ell=1}^a
 t(\muu_k,\la_\ell)\right)  +\frac{2c^2}{\muu_{jk}^2-c^2},
 \end{equation}
where $v_{km}=v_k-v_m$ and $j,k=1,\dots,b$. The antidiagonal blocks have more simple structure
 \begin{equation}\label{P12}
\Not_{j,a+k}=t(\muu_k,\la_j),\qquad j=1,\dots,a,\quad k=1,\dots,b,
 \end{equation}
 \begin{equation}\label{P21}
\Not_{a+j,k}=t(\muu_j,\la_k),\qquad j=1,\dots,b,\quad k=1,\dots,a.
 \end{equation}
Observe that the matrix $\Not$ is symmetric:
$\Not_{jk}=\Not_{kj}$.
It is also easy to check (see \cite{BelPRS12b}) that
\be{matN-norm}
\begin{aligned}
&\Not_{j,k}=-c\frac{\partial\Phi_j}{\partial u_k}, \qquad &j=1,\dots,a+b,\quad k=1,\dots,a;
\num
&\Not_{j,a+k}=c\frac{\partial\Phi_{j}}{\partial v_k},  &j=1,\dots,a+b,\quad  k=1,\dots,b,
\end{aligned}
\ee
where $\Phi_j$ is given by \eqref{Phi-1}, \eqref{Phi-2}.

Let us reproduce the result \eqref{Qm-2} starting form the representation \eqref{SP-res}. The entries
of the matrix $\Not^{(s)}_{j,k}$ with $j,k=1,\dots,a+b$ coincide with the ones defined in
\eqref{P11}--\eqref{P21}. In the last row we have
\be{N-Y2}
\begin{aligned}
&\Not^{(s)}_{a+b+1,k}=Y^{(s)}_k=\delta_{s2}-\delta_{s1},\qquad k=1,\dots,a;\num
&\Not^{(s)}_{a+b+1,k}=Y^{(s)}_{k}=\delta_{s2}-\delta_{s3},\qquad k=a+1,\dots,b.
\end{aligned}
\ee
Finally, the last column has the components
\be{last-column}
\begin{aligned}
&\Not^{(s)}_{j,a+b+1}=\frac{c}{f(z,\bla)f(\bmu,z)}\,\frac{\partial\tau(z|\bla,\bmu)}{\partial u_j}, \qquad &j=1,\dots,a;\num
&\Not^{(s)}_{a+j,a+b+1}=-\frac{c}{f(z,\bla)f(\bmu,z)}\,\frac{\partial\tau(z|\bla,\bmu)}{\partial v_j}, \qquad &j=1,\dots,b\\
&\Not^{(s)}_{a+b+1,a+b+1}=\frac{1}{f(z,\bla)f(\bmu,z)}\,\frac{\partial\tau_{\bar\kappa}(z|\blac,\bmuc)}{\partial\kappa_s}
\Bigr|_{\bar\kappa=1}.
\end{aligned}
\ee

Thus, we have described the $(a+b+1)\times(a+b+1)$ matrix $\Not^{(s)}$ in the limit $\blac=\blab=\bla$ and $\bmuc=\bmub=\bmu$. Let us show that $\det\Not^{(s)}$ can be reduced to the determinant of the $(a+b)\times(a+b)$ block of this matrix given by \eqref{P11}--\eqref{P21}. For this we introduce three $(a+b)$-component vectors $\tilde\Omega^{(s)}_j$ as
\be{def-tOmega}
\begin{array}{cl}
{\dis \tilde\Omega^{(s)}_j=\frac1c\,\frac{d\lac_j}{d\kappa_s}\Bigr|_{\bar\kappa=1}
,\qquad}&{\dis j=1,\dots,a,}\num
{\dis \tilde\Omega^{(s)}_{a+j}=-\frac{1}c\,\frac{d\muc_j}{d\kappa_s}\Bigr|_{\bar\kappa=1},\qquad}&
{\dis j=1,\dots,b.}
\end{array}
\ee
It is easy to show that
\be{Sum-N}
\sum_{j=1}^{a+b+1}\tilde\Omega^{(s)}_j\Not^{(s)}_{j,k}=0,\qquad k=1,\dots,a+b.
\ee
Indeed, differentiating the system of twisted Bethe equations \eqref{Log-TBE} with respect to
$\kappa_s$ at $\bar\kappa=1$ we obtain
\be{diff-Log-TBE}
\sum_{\ell=1}^a\frac{\partial\Phi_j}{\partial u_\ell}\frac{d\lac_\ell}{d\kappa_s}\Bigr|_{\bar\kappa=1}+
\sum_{m=1}^b\frac{\partial\Phi_j}{\partial v_m}\frac{d\muc_m}{d\kappa_s}\Bigr|_{\bar\kappa=1}
=Y_k^{(s)}.
\ee
Taking into account \eqref{matN-norm} and the symmetry of the matrix $\Not^{(s)}_{j,k}$ for $j,k=1,\dots,a+b$
we immediately arrive at \eqref{Sum-N}. Thus, adding to the last row of the matrix $\Not^{(s)}_{j,k}$ all other
rows multiplied by the coefficients $\tilde\Omega^{(s)}_j$ we obtain zeros everywhere except the element $j,k=a+b+1$,
where we have
\begin{multline}\label{last-elem}
\sum_{j=1}^{a+b+1}\tilde\Omega^{(s)}_j\Not^{(s)}_{j,a+b+1}=
\frac{1}{f(z,\bla)f(\bmu,z)}\biggl\{\frac{\partial\tau(z|\blac,\bmuc)}{\partial\kappa_s}
\Bigr|_{\bar\kappa=1}\\
+\sum_{\ell=1}^a\frac{\partial\tau(z|\bla,\bmu)}{\partial u_\ell}
\frac{d\lac_\ell}{d\kappa_s}\Bigr|_{\bar\kappa=1}+
\sum_{m=1}^b\frac{\partial\tau(z|\bla,\bmu)}{\partial v_m}\frac{d\muc_m}{d\kappa_s}\Bigr|_{\bar\kappa=1}
\biggr\}=\frac{d\tau(z|\blac,\bmuc)}{d\kappa_s}
\Bigr|_{\bar\kappa=1}\;.
\end{multline}
Thus, we obtain
 \be{SP-res-1}
 \mathcal{F}_{a,b}^{(s,s)}(z|\bla,\bmu;\bla,\bmu)=
 \frac{d\tau(z|\bla,\bmu)}{d\kappa_s}\Bigr|_{\bar\kappa=1}\cdot H_{a,b}\det_{a+b}\Not.
 \ee
Comparing \eqref{SP-res-1} with \eqref{Norm} we arrive at the representation \eqref{Qm-2}.

\section{Discussions\label{S-Disc}}

In this paper we considered the form factors of the monodromy matrix entries in the models with
$GL(3)$-invariant $R$-matrix  and obtained determinant representations for them.
The question arises of generalizing the results obtained to the models with the symmetry group of a higher rank.
For this it is useful to compare the structure of the determinant formulas for the models with $GL(2)$ and
$GL(3)$ symmetry.

For $GL(3)$-based models
all the representations have similar structure and
are based on the determinants of the matrix $\Not$ or $\Ntt$ (the last one can be obtained from
$\Not$ by the replacement $\{\blac,\bmuc\}\leftrightarrow \{\blab,\bmub\}$). In these matrices
all rows and columns  are associated with one of the Bethe parameters or with the external variable $z$.
Say, in the matrix $\Not$ the first $a$ columns  correspond to the set $\blab$, the next $b$
columns correspond to the set $\bmuc$, and the last column is associated with the variable $z$.
The rows of this matrix are associated with the parameters $\blac$, $\bmub$.
For the form factor of the diagonal entries,  as well as for the operator $T_{13}(z)$,  the matrix $\Not$ has
an additional row.

It is hardly possible to predict such the structure based on the results obtained for
the models possessing $GL(2)$ symmetry. One could expect that, for example, the columns of the matrices should correspond
to the parameters of one Bethe vector (say, $\{\blab,\bmub\}$), while the rows should correspond
to the parameters of another Bethe vector (in this case, $\{\blac,\bmuc\}$). We see, however, that it is not the case,
and one should `mix' the parameters from different Bethe vectors in order to label the rows
and the columns.

Such mixing of the Bethe parameters makes very problematic a  straightforward generalization of our results
to the models with $GL(N)$-symmetry with $N>3$.
There exists also one more argument to rule out a simple generalization of these results to the symmetry groups
of higher rank. We see that the matrix whose determinant describe form factors, have a block structure
\be{block-matrix}
\Not=\begin{pmatrix}
\quad\Not_\ell\quad\\
-\;-\;-\;-\\
\quad\Not_r\quad
\end{pmatrix},\qquad\text{where}\qquad
\begin{array}{l}
(\Not_\ell)_{j,k}\sim\frac{\partial\tau(x_k|\blac,\bmuc)}{\partial\lac_j},\\
(\Not_r)_{j,k}\sim\frac{\partial\tau(x_k|\blab,\bmub)}{\partial\mub_j}.
\end{array}
\ee
The upper and lower blocks are proportional to the Jacobians
of the transfer matrix eigenvalues on the left and the right Bethe vectors respectively. On the other hand the
block structure is also related to the fact that Bethe vectors depend on two sets of parameters.
However, in the case of the $GL(N)$ group, Bethe vectors
depend on $N-1$ sets of variables \cite{KulRes83}. Hence, it is natural to expect that if there are determinant representations for form factors in the models with symmetry group, for example
$GL(4)$, then the corresponding matrices should have a block structure $3\times 3$. At the same time we still have
only two vectors and, hence, only two eigenvalues.

Of course, the arguments above do not mean that determinant representations for form factors do not exist
in the models with $GL(N)$-invariant $R$-matrix. These arguments can only tell that the determinant representations based on the Jacobians of the  transfer matrix eigenvalues are hardly possible for models with higher symmetry group.
However, on the other hand, we can not exclude the existence of determinant representations having different structure.

Concluding this paper we would like to say few words about possible applications. One of them immediately
arises for the quantum  models admitting explicit solution of the quantum inverse scattering problem
\cite{KitMT99,MaiT00}. In particular, one has the following representation for the local operators
in the $SU(3)$-invariant XXX Heisenberg
chain:
 \be{gen-sol-T}
 E^{\alpha,\beta}_m =(\tr T(0))^{m-1}  T_{\beta\alpha}(0)(\tr T(0))^{-m}.
 \ee
Here $E^{\alpha,\beta}_m$,
$\alpha,\beta=1,2,3$, is an elementary
unit  ($\left(E^{\alpha,\beta}\right)_{jk}=\delta_{j\alpha}\delta_{k\beta}$)
associated with the $m$-th site of the chain. Since the action of the transfer matrix
$\tr T(0)$ on on-shell Bethe vectors  is trivial, we see that the form factors of $E^{\alpha,\beta}_m$
are proportional to those of $T_{\beta\alpha}$
 \be{ffE-ffT}
 \mathbb{C}^{a',b'}(\blac;\bmuc)E^{\alpha,\beta}_m\mathbb{B}^{a,b}(\blab;\bmub)
 =\frac{\tau^{m-1}(0|\blac,\bmuc)}{\tau^{m}(0|\blab,\bmub)}
 \mathcal{F}_{a,b}^{(\beta,\alpha)}(0|\blac,\bmuc;\blab,\bmub).
 \ee
Thus, if we have an explicit and compact representations
for form factors of $T_{\beta,\alpha}$, we can study the problem of two-point and multi-point correlation functions, expanding them into series with respect to the form factors.

\section*{Acknowledgements}
The work of S.P. was supported in part by RFBR grant 14-01-00474-a and  grant
of Scientific Foundation of NRU HSE. E.R. was supported by ANR Project
DIADEMS (Programme Blanc ANR SIMI1 2010-BLAN-0120-02).
N.A.S. was  supported by the Program of RAS Basic Problems of the Nonlinear Dynamics,
RFBR-11-01-00440-a, RFBR-13-01-12405-ofi-m2.

\appendix

\section{Summation formulas}

In this section we prove several identities for the vector $\Omega$ introduced in \eqref{def-Omega}.
\begin{prop}\label{Omega-sum}
Let $\Omega$ is defined as in \eqref{def-Omega}. Then
\be{tuz}
\begin{aligned}
\sum_{j=1}^at(\lac_j,z)\Omega_j&=\frac{h(\blab,z)}{h(\blac,z)}\left(1-\frac{f(\blac,z)}{f(\blab,z)}\right),\\
\sum_{j=1}^at(z,\lac_j)\Omega_j&=\frac{h(z,\blab)}{h(z,\blac)}\left(\frac{f(z,\blac)}{f(z,\blab)}-1\right),\\
\sum_{j=1}^bt(\mub_j,z)\Omega_{j+a}&=\frac{h(\bmuc,z)}{h(\bmub,z)}\left(1-\frac{f(\bmub,z)}{f(\bmuc,z)}\right),\\
\sum_{j=1}^bt(z,\mub_j)\Omega_{j+a}&=\frac{h(z,\bmuc)}{h(z,\bmub)}\left(\frac{f(z,\bmub)}{f(z,\bmuc)}-1\right).
\end{aligned}
\ee
\end{prop}
All the identities above can be proved in a similar way. Consider, for example, the first identity.

{\sl Proof}. Let
 \be{sim-sum}
 \sum_{j=1}^at(\lac_j,z)\Omega_j=W(z).
\ee
The sum in the l.h.s. of \eqref{sim-sum} can be computed by means of an auxiliary integral
\be{Integ}
I=\frac1{2\pi i}\oint\limits_{|\omega|=R\to\infty}
\frac{c\,d\omega}{(\omega-z)(\omega-z+ c)}
\prod\limits_{\ell=1}^a \frac{\omega-\lab_\ell}{\omega-\lac_\ell}.
\ee
The integral is taken over the anticlockwise oriented contour $|\omega|=R$ and we consider the limit $R\to\infty$. Then $I=0$, because the integrand behaves as $1/\omega^2$ at $\omega\to\infty$. On the other hand the same integral is equal to the sum
of the residues within the integration contour. Obviously the sum of the residues at $\omega=\lac_\ell$ gives
$W(z)$. There are also two additional poles at $\omega=z$ and $\omega=z-c$. Then we have
 \be{I-res}
 I=0=W(z)- \prod_{\ell=1}^a\frac{z-\lab_\ell- c}{z-\lac_\ell- c}+\prod_{\ell=1}^a\frac{z-\lab_\ell}{z-\lac_\ell}.
 \ee
From this we obtain the first identity \eqref{tuz}


\begin{thebibliography}{99}

\bibitem{KarW78}
M. Karowski and P. Weisz, {\sl Exact form factors in $(1 + 1)$-dimensional field theoretic models with soliton behaviour},
Nucl. Phys. B {\bf 139} (1978) 455--476.
\bibitem{Smi92b}
F. A. Smirnov,
{\sl Form factors in completely integrable models of quantum field theory},  Adv. Series in Math. Phys.  {\bf 14}, World Scientific, 1992.
%
\bibitem{CarM90}
J. Cardy and G. Mussardo, {\sl Form factors of descendent operators in perturbed conformal field theories},
Nucl. Phys. B {\bf 340} (1990) 387--402.
\bibitem{Mus92}
G. Mussardo, {\sl Off-critical statistical models: Factorized scattering theories and bootstrap program},
Phys. Rep. {\bf 218} (1992) 215--379.
\bibitem{FriMS90} A. Fring, G. Mussardo and P. Simonetti, {\sl Form factors for integrable lagrangian field theories,
the sinh-Gordon model}, Nucl. Phys. B {\bf 393} (1990) 413--441, \texttt{arXiv:hep-th/9211053}.
\bibitem{KouM93} A. Koubek and G. Mussardo, {\sl On the operator content of the sinh-Gordon model},
Phys. Lett. B {\bf 311} (1993) 193--201, \texttt{arXiv:hep-th/9306044}.
\bibitem{AhnDM93} C. Ahn, G. Delfino and G. Mussardo, {\sl Mapping between the sinh-Gordon and Ising models},
Phys. Lett. B {\bf 317} (1993) 573--580, \texttt{arXiv:hep-th/9306103}.
\bibitem{Zam91}
A. B. Zamolodchikov, {\sl Two-point correlation function in scaling Lee-Yang model},
Nucl. Phys. B {\bf 348} (1991) 619--641.
\bibitem{LukZ97}
S. Lukyanov and A.  Zamolodchikov, {\sl Exact expectation values of local fields in the quantum sine-Gordon model},
Nucl. Phys. B {\bf 493} (1997) 571--587, \texttt{arXiv:hep-th/9611238}.
\bibitem{Luk99}
S.~Lukyanov, {\sl Correlation amplitude for the $XXZ$ spin chain in the disordered regime},
Phys. Rev. B {\bf 59} (1999) 11163--11164,  \texttt{arXiv:cond-mat/9809254}.
\bibitem{LukZ01}
S. Lukyanov and A.  Zamolodchikov, {\sl Form factors of soliton-creating operators in the sine-Gordon model},
Nucl. Phys. B {\bf 607} (2001) 437--455, \texttt{arXiv:hep-th/0102079}.
\bibitem{JimMMN92}
M.~Jimbo, K.~Miki, T.~Miwa and A.~Nakayashiki, {\sl Correlation functions of the XXZ model for
$\Delta<-1$}, Phys. Lett. A {\bf 168} (1992) 256--263, \texttt{arXiv:hep-th/9205055}.
\bibitem{JimM95L}
M.~Jimbo and T.~Miwa,
{\sl Algebraic analysis of solvable lattice models}, Regional Conference Series in Mathematics, vol 85,  AMS, 1995.
\bibitem{JimM96}
M.~Jimbo and T.~Miwa, {\sl Quantum KZ equation with $|q| = 1$ and correlation functions of the
$XXZ$ model in the gapless regime},
J. Phys. A: Math. Gen. {\bf 29}  (1996) 2923--2958, \texttt{arXiv:hep-th/9601135}.
%
\bibitem{KojKS97}
T. Kojima, V. Korepin, N. Slavnov, {\sl Determinant representation for dynamical correlation functions of the Quantum nonlinear Schr\"odinger equation},  Commun. Math. Phys. {\bf 188} (1997) 657--689, \texttt{arXiv:hep-th/9611216}.
\bibitem{KitMT99} N. Kitanine, J. M. Maillet and V. Terras, {\sl Form factors of the XXZ Heisenberg spin-12 finite chain},
Nucl. Phys. B  {\bf 554}  (1999) 647--678, \texttt{arXiv:math-ph/9807020}.
%
\bibitem{FadST79} L. D. Faddeev, E. K. Sklyanin and L. A. Takhtajan, {\sl Quantum Inverse Problem. I},
 Theor. Math. Phys. {\bf 40} (1979) 688--706.
%
\bibitem{FadT79} L. D. Faddeev and L. A. Takhtajan, {\sl The quantum method of the inverse problem and the Heisenberg $XYZ$ model},
Usp. Math. Nauk {\bf 34} (1979) 13;  Russian Math. Surveys {\bf 34} (1979) 11 (Engl. transl.).
%
\bibitem{BogIK93L}V. E. Korepin, N. M. Bogoliubov,
A. G. Izergin, {\sl Quantum Inverse Scattering Method and Correlation Functions}, Cambridge: Cambridge Univ.
Press, 1993.
%
\bibitem{FadLH96} L. D. Faddeev, in: Les Houches Lectures {\sl Quantum Symmetries}, eds A. Connes
et al, North Holland, (1998) 149, \texttt{arXiv:hep-th/9605187}.
\bibitem{MaiT00} J.M. Maillet, V. Terras,{\sl On the quantum inverse scattering problem},
Nucl. Phys. B {\bf 575} (2000) 627--644, \texttt{arXiv:hep-th/9911030}.
%
\bibitem{KitKMST11} N. Kitanine, K. Kozlowski, J. M. Maillet, N. A. Slavnov, V. Terras, {\sl
A form factor approach to the asymptotic behavior of correlation functions}, J. Stat. Mech. (2011) P12010,
\texttt{arXiv:hep-th/1110.0803}.
%
\bibitem{KitKMST12} N. Kitanine, K. Kozlowski, J. M. Maillet, N. A. Slavnov, V. Terras, {\sl
Form factor approach to dynamical correlation functions in critical models}, J. Stat. Mech. (2012) P09001,
\texttt{arXiv:1206.2630}.
%
\bibitem{CauM05}
J.~S. Caux and J.~M. Maillet, {\sl Computation of Dynamical Correlation Functions of
Heisenberg Chains in a Magnetic Field},
Phys. Rev. Lett. {\bf 95} (2005) 077201 3pp, \texttt{arXiv:cond-mat/0502365}.
%
\bibitem{CauPS07}J.~S. Caux, P. Calabrese, N.~A. Slavnov, {\sl One-particle dynamical correlations in the
one-dimensional Bose gas},  J. Stat. Mech. (2007) P01008, \texttt{arXiv:cond-mat/0611321}.
%
\bibitem{su4-Hub} R. Assaraf, P. Azaria, E. Boulat, M. Caffarel, P. Lecheminant, \textsl{Dynamical Symmetry Enlargement Versus Spin-Charge Decoupling in the One-Dimensional SU(4) Hubbard Model}, Phys. Rev. Lett. \textbf{93} (2004) 016407, \texttt{arXiv:cond-mat/0310090}.
%
\bibitem{so8-Hub} H.-H. Lin, L. Balents and M. P. A. Fisher, \textsl{Exact SO(8) symmetry in the weakly-interacting two-leg ladder}, Phys. Rev. B \textbf{58} (1998) 1794╨1825.
%
\bibitem{PozOK12} B. Pozsgay, W.-V. van G. Oei and M. Kormos, {\sl On Form Factors in nested Bethe Ansatz systems},
J. Phys. A: Math. Gen. {\bf 45}  (2012) 465007, \texttt{arXiv:1204.4037}
%
 \bibitem{KulRes83}
P. P. Kulish, N. Yu. Reshetikhin,
{\sl Diagonalization of $GL(N)$ invariant transfer matrices and quantum $N$-wave system (Lee model)}, J.~Phys.~A:  {\bf 16} (1983) L591--L596.
%
 \bibitem{KulRes81}
P. P. Kulish, N. Yu. Reshetikhin,
{\sl Generalized Heisenberg ferromagnet and the Gross--Neveu model}, Zh. Eksp. Theor. Fiz.
{\bf 80} (1981) 214--228; Sov. Phys. JETP,  {\bf 53}:1 (1981)  108--114 (Engl. transl.)
%
 \bibitem{KulRes82}
P. P. Kulish, N. Yu. Reshetikhin,
{\sl GL(3)-invariant solutions of the Yang-Baxter equation and associated quantum systems}, Zap. Nauchn. Sem. POMI.
{\bf 120} (1982) 92--121; J. Sov. Math.,  {\bf 34}:5 (1982)  1948--1971 (Engl. transl.)
%
\bibitem{BelPRS12b} S. Belliard, S. Pakuliak, E. Ragoucy, N. A. Slavnov,
{\sl The algebraic Bethe ansatz for scalar products in $SU(3)$-invariant integrable
 models}, J. Stat. Mech. (2012) P10017, \texttt{arXiv:1207.0956}.
%
\bibitem{BelPRS13a} S. Belliard, S. Pakuliak, E. Ragoucy, N. A. Slavnov,
{\sl Form factors in  $SU(3)$-invariant integrable models}, J. Stat. Mech.  (2013) P04033, \texttt{arXiv:1211.3968}.
%
%
\bibitem{PakRS14b}
    S. Pakuliak, E. Ragoucy, N. A. Slavnov,
\textsl{Form factors in quantum integrable models with GL(3)-invariant R-matrixФ,}
Nucl. Phys. B,  {\bf 881} (2014) 343Ц368, \texttt{arXiv:1312.1488}.
%
\bibitem{BelPRS12c} S. Belliard, S. Pakuliak, E. Ragoucy, N. A. Slavnov,
{\sl Bethe vectors of $GL(3)$-invariant integrable models}, J. Stat. Mech. (2013) P02020, \texttt{arXiv:1210.0768}.
%
\bibitem{Sla89} N. A. Slavnov, {\sl Calculation of scalar products of wave functions and form factors in the framework of the algebraic Bethe ansatz}, Theor. Math. Phys. {\bf 79}:2 (1989) 502--508.
%
    \bibitem{KitKMST05b} N. Kitanine, K.~K. Kozlowski, J. M. Maillet, N. A. Slavnov and V. Terras,
{\sl  On correlation functions of integrable models associated to the six-vertex R-matrix},
J. Stat. Mech. 0701 (2007) P01022, \texttt{arXiv:hep-th/0611142}.
%
\bibitem{Res86}  N. Yu. Reshetikhin, {\sl Calculation of the norm of Bethe vectors in models with $SU(3)$-symmetry}, Zap. Nauchn. Sem. LOMI {\bf 150} (1986) 196--213;    J. Math. Sci. {\bf 46} (1989) 1694--1706 (Engl. transl.).
%
\bibitem{Whe12} M. Wheeler, {\sl Scalar products in generalized models with $SU(3)$-symmetry},
Comm. Math. Phys. \textbf{327} (2014) 737-777, \texttt{arXiv:1204.2089}.
%
\bibitem{BelPRS12a} S. Belliard, S. Pakuliak, E. Ragoucy, N. A. Slavnov,
{\sl Highest coefficient of scalar products in $SU(3)$-invariant models}, J. Stat. Mech.  (2012) P09003, \texttt{arXiv:1206.4931}
%
\bibitem{Kor82} V. E. Korepin, {\sl Calculation of norms of Bethe wave functions}, Commun. Math. Phys.
{\bf 86} (1982) 391--418.
%
\bibitem{IzeK84} A. G. Izergin,  V. E. Korepin,
{\sl The quantum inverse scattering method approach to correlation functions},
Commun. Math. Phys. {\bf 94} (1984), 67--92.
%
\bibitem{KitMST05} N. Kitanine, J. M. Maillet, N. A. Slavnov and V. Terras,
{\sl Master equation for spin-spin correlation functions of the
$XXZ$ chain}, Nucl. Phys. B {\bf 712} (2005) 600, \texttt{arXiv:hep-th/0406190}.
%
\end{thebibliography}
\end{document}